\documentclass[12pt,preprint]{aastex}
\usepackage{natbib}
\usepackage{amsmath}    

\newcommand{\fest}{\mbox{FeSt~1-457}}
\newcommand{\coI}{\mbox{C$^{18}$O ($J = 1 \to 0$)}}
\newcommand{\coII}{\mbox{C$^{18}$O ($J = 2 \to 1$)}}
\newcommand{\co}{C$^{18}$O}
\newcommand{\hcoI}{\mbox{HCO$^+$ ($J = 1 \to 0$)}}
\newcommand{\hco}{HCO$^+$}
\newcommand{\csII}{\mbox{CS ($J = 2 \to 1$)}}

\newcommand{\nhI}{\mbox{N$_2$H$^+$ ($J = 1 \to 0$)}}
\newcommand{\nh}{N$_2$H$^+$}
\newcommand{\ud}{\mathrm{d}}

\slugcomment{To appear in ApJ 20 August 2007}

\shorttitle{Dynamical State of FeSt 1-457}
\shortauthors{Aguti et al.}

\begin{document}

\title{The Dynamical State of the Starless Dense Core FeSt 1-457:
A Pulsating Globule?}

\author{E.~D. Aguti\altaffilmark{1,2}, C.~J. Lada\altaffilmark{2},
  E.~A. Bergin\altaffilmark{3}, J.~F Alves\altaffilmark{4}, and
 M. Birkinshaw\altaffilmark{1}}

\altaffiltext{1}{Astrophysics Group, University of Bristol,  Tyndall
Avenue, Bristol, BS8 1TL, UK. \mbox{esther.aguti@bristol.ac.uk}}
\altaffiltext{2}{Harvard-Smithsonian Center for Astrophysics, 60 Garden Street,
Cambridge, MA 02138, USA. \mbox{clada@cfa.harvard.edu}}
\altaffiltext{3}{Department of Astronomy, University of Michigan, 825
Dennison Building, 501 East University Avenue, Ann Arbor, MI 48109,
USA. \mbox{ebergin@umich.edu}}
\altaffiltext{4}{Calar Alto Observatory, Centro Astron\'{o}mico Hispano
Alem\'{a}n, c/Jes\'{u}s Durb\'{a}n Rem\'{o}n 2-2, 04004
Almeria, Spain. \mbox{jalves@caha.es}}


\begin{abstract}

High resolution molecular line observations of \csII, \hcoI, \coI, \coII\ and
\nhI\ were obtained toward the starless globule \fest\ in order to investigate
its kinematics and chemistry.  The HCO$^+$ and CS spectra show clear self-reversed
and asymmetric profiles across the face of the globule.  The sense of the
observed asymmetry is indicative of the global presence of expansion motions in
the outer layers of the globule.  These motions appear to be
subsonic and significantly below the escape velocity of the globule.  Comparison
of our observations with near-infrared extinction data indicate that the globule
is gravitationally bound.  Taken together these considerations lead us to
suggest that the observed expansion has its origin in an oscillatory motion of
the outer layers of the globule which itself is likely in a quasi-stable state
near hydrostatic equilibrium.  Analysis of the observed linewidths of \co\ and
\nhI\ confirm that thermal pressure is the dominant component of the cloud's
internal support.  A simple calculation suggests that the dominant mode of
pulsation would be an l = 2 mode with a period of $\sim$ $3 \times 10^5$ yrs.
Deformation of the globule due to the large amplitude l = 2 oscillation may be
responsible for the double-peaked structure of the core detected in high
resolution extinction maps.

Detailed comparison of the molecular-line observations and extinction data
provides evidence for significant depletion of \co\ and perhaps \hco\ while
\nhI\ may be undepleted to a cloud depth of $\sim$ 40 magnitudes of 
visual extinction.

\end{abstract}


\keywords{ISM:clouds, ISM: globules, ISM: molecules, ISM: 
individual(\object[FEST 1-457]{FeSt 1-457})}

\section{Introduction}

A key component to understanding star formation lies in determining the
dynamical nature of globules and cores of molecular clouds from which stars
form.  This is typically done using molecular lines, and observations of
asymmetry or self-reversal in the line profiles of optically thick species
portray the motion of the outer layers of material within a cloud.  Many of the
spectra measured from starless cores and globules have blue asymmetirc profiles,
such that the blue side of the profile is stronger than the red side, suggestive
of red-shifted self-absorption and infall motion.  For example, in their survey
of starless cores, \citet{lee01} found that approximately one third of the cores
they observed displayed bona fide infall motions in the outer layers and
therefore were consistent with expectations of collapse.  In this survey and in
a number of others, however, there were a number of cores that displayed
evidence for blue-shifted self-absorption in their spectra, e.g., CB246, L1521F
and L429-1.  This is an indication of expansion but with no protostar in the
core, what mechanism drives the expanding motions?  There are also cores and
globules which show both redshifted and blueshifted self-absorption
simultaneously in a single map of one molecular transition, for example, L1512,
L1689B, L1544 \citep{lee01}, and B68 \citep{lada03}.  Other cores and globules
have spectra that shift from redshifted to blueshifted self-absorption depending
on which molecular transition is observed.  For example, self -absorption in the
source L1157 is redshifted in the HCO$^+$($J = 3 \to 2$) line profile but
blueshifted in the HCO$^+$($J = 4 \to 3$) line profile \citep{greg97}.  Clearly
complex motions are occurring in the various layers of the cores but it is not
clear what causes them.

\citet{redm04} concluded that the pattern of redshifted and
blueshifted self-absorption observed across the source in L1689 is caused by
rotational motion in the globule.  In B68, where the spatial pattern
switches from blueshifted to redshifted profiles both east and west of
the center of the globule, the observed line profiles cannot be
explained by rotation.  \citet{lada03} proposed that B68 could be in a
quasi-stable state near hydrostatic equilibrium, and the systematic
velocity field observed is likely to be caused by oscillatory motions
within the outer layers of the globule.  For stable and quasi-stable
clouds, such motions are predicted by theory, e.g., \citet{mats03}
showed that small amounts of rotation can cause an initially
gravitationally unstable cloud to stabilize and oscillate;
\citet{clar97} in considering the thermal and dynamical balance in low
mass dense cores find that a quasi-equilibrium state can be
established, which is not static but is pulsationally
stable, and in their hydrodynamic study of starless cores,
\citet{keto05} found that cores could oscillate with periods of about
one million years (or the sound crossing time) if perturbed by a
modest external force.  Such oscillatory behavior is a signature of
dynamical equilibrium, and therefore, the presence of such complex
patterns in their spectral line maps could indicate pulsationally
stable initial states for the star-forming cores and globules.

In this study, we have obtained high angular resolution spectra of the
source listed in the catalogue of \citet{feit84} as \fest.  It is a
small, round, dark globule situated in the Pipe Nebula with no
associated IRAS, MSX or Spitzer Space Telescope point sources, and is
therefore almost certainly starless.  Its distance has not been
directly determined but it is associated with the Pipe Nebula, for which
some distance estimates have been made. The most reliable estimate to
date is 130~pc determined by using infrared extinction measurements
for Hipparcos stars, \citep{lomb06}.  In their Bonnor-Ebert sphere
fitting analysis, however, \citet{kand05} report a distance of 70~pc,
which compares to the value of 90~pc derived in a similar fashion for
B68, also, incidently, part of the Pipe Nebula complex.  In this
paper, we adopt the \citet{lomb06} distance of 130 pc.  The angular
diameter of \fest, as apparent in optical images, is $\sim$
4.5$\arcmin$, which is $\sim$ 35,000 AU (0.17 pc) at 130 pc.  Section
\ref{sec:obs} describes our observations of \fest, and
\S\ref{sec:results}, is a description of the results.  We detected
expanding motions in \hco and CS line profiles across the surface of
\fest\ but since no protostar has been detected, the source of this
expansion is a mystery.  In \S\ref{sec:disc}, we show that the core is
bound and self-gravitating and we propose that pulsation or
oscillatory motion within the cloud layers may account for the
observed expanding motions.  Finally, we present evidence for
molecular depletion in the central regions of the core.

\section{Observations}\label{sec:obs}

The observations reported here were made in July 2003 using the 30-m
IRAM millimeter-wave telescope at Pico Veleta in Spain.  The dual
mixer, dual-channel receiver was tuned to observe the molecules listed
with their respective frequencies, beamwidths and velocity resolutions
in Table~\ref{tb:linpars}.  The frequency-switching mode was used to
obtain the observations, and system temperatures were 150 - 200~K.
The chopper wheel method was used for calibration.  The spectrometer
is an autocorrelator configured to give the velocity resolutions
listed in Table~\ref{tb:linpars}.  Beam efficiencies were $\sim$ 0.75
for the lower frequency 80 - 115~GHz range, and $\sim$ 0.6 for the
higher frequency 197 - 266~GHz range.  Observations were made in a
grid with a reference position at \mbox{$\alpha_{2000}$ =
17$^h$35$^m$47$\fs$5}, \mbox{$\delta_{2000}$ =
-25$\degr$33$\arcmin$2$\farcs$0}.  \hco, \coI\ and \coII\ were
uniformly sampled such that a region centered on the reference
position was observed every 24$\arcsec$ out to a maximum distance of
96 $\arcsec$ in Right Ascension and 120$\arcsec$ in Declination, a
total of 9 $\times$ 11 or 99 spectra for each species.  This grid spacing is
approximately one beamwidth, slightly over for \hco\ and under for
\co.  The \nh\ and CS emission was much weaker and less extended so
the grid size was reduced to 5 $\times$ 5 around the reference point
with an additional two measurements directly north, east, south and
west of the center, resulting in a total of 33 spectra for these
species.  The spacing between measurements was also 24$\arcsec$,
approximately one beamwidth for each of the molecules.  Data reduction
was done using the standard IRAM package CLASS and IDL (Interactive
Data Language by Research Systems, Inc.).

\section{Results}\label{sec:results}

\subsection{Self-Reversed HCO$^+$ and CS Lines} \label{sec:hco}

\begin{figure}
\begin{center}
\includegraphics[height=4in]{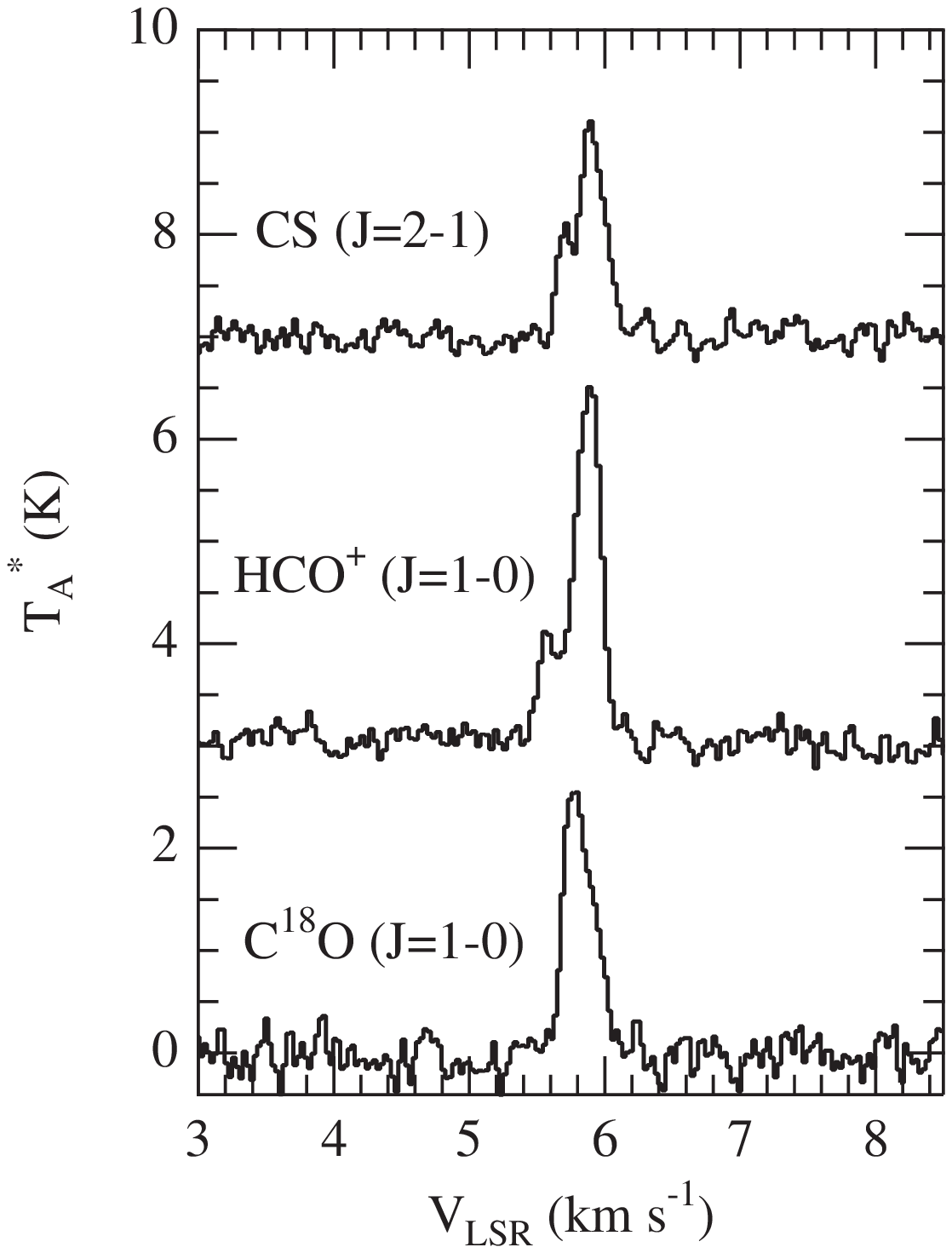}
\caption{Emission-line spectra of \csII, \hcoI\ and \coI\
near the center of \fest\ measured using the 30m
IRAM telescope.  Split asymmetric, double-peaked profile
shapes characterize the CS and HCO$^+$ lines but not the 
C$^{18}$O line which is single-peaked indicating that the former
emission lines are likely very optically thick and self-reversed. The sense
of the asymmetry suggests expansion motions of the outer cloud layers.}
\label{spec}
\end{center}
\end{figure}

\begin{figure}
\begin{center}
\includegraphics[height=4in]{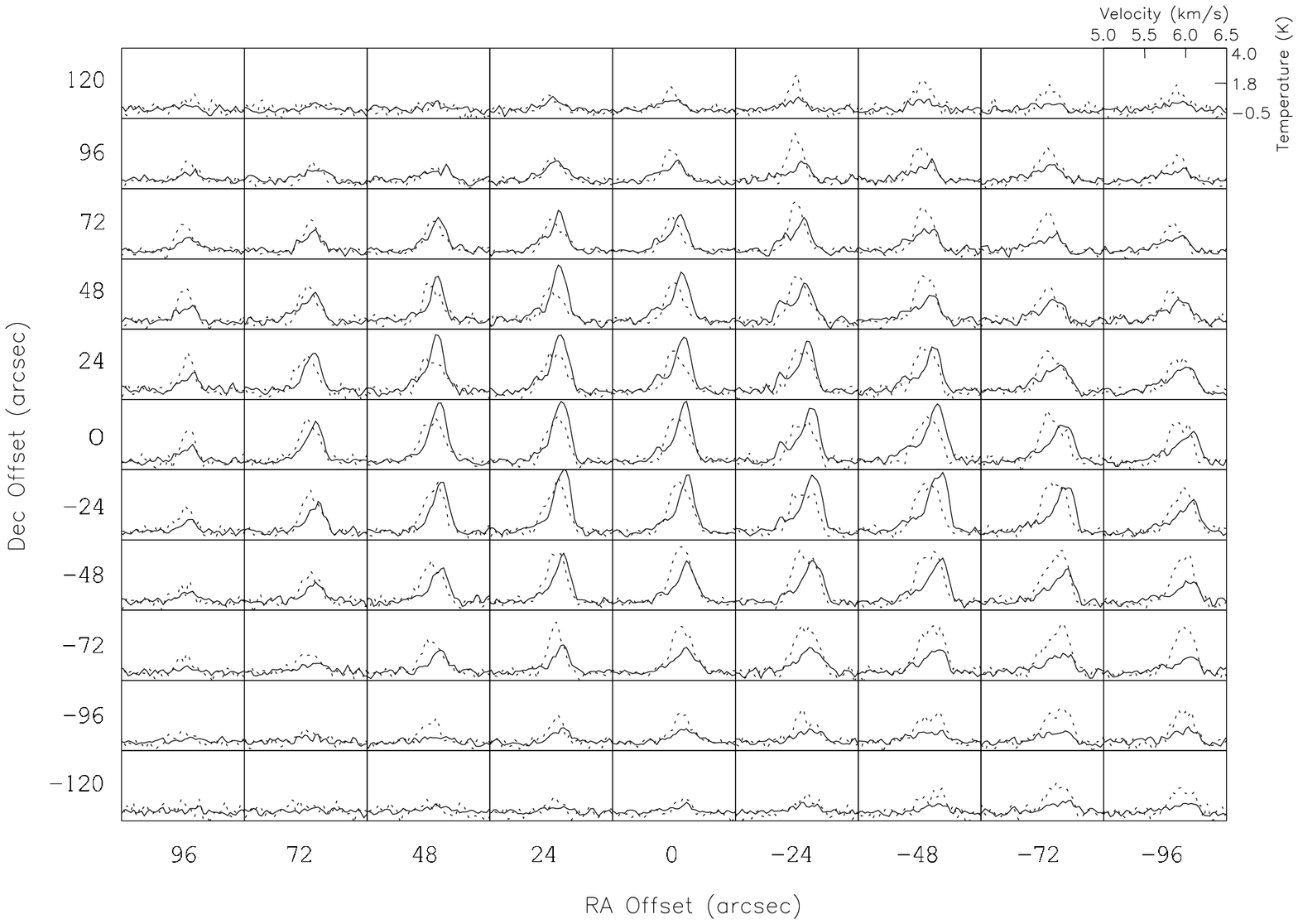}
\caption{Map of emission-line spectra from across \fest.  
Solid lines are HCO$^+$ emission spectra and dotted
lines are those of C$^{18}$O.  The map is centered at
$\alpha_{2000}$ = 17$^h$35$^m$47.5$^s$, $\delta_{2000}$ =
-25$\degr$33$\arcmin$2.0$\arcsec$.  Double-peaked, self-reversed 
and asymmetric profiles 
are evident in
the HCO$^+$ lines across the globule. In all positions the sense of
the profile asymmetry is indicative of expansion motions.}
\label{c18ohcomap}
\end{center}
\end{figure}

Figure \ref{spec} shows the \csII, \hcoI\ and \coI\ spectra at the reference
position.  The CS and HCO$^+$ profiles display a split asymmetric, double-peaked
shape while the C$^{18}$O line is single-peaked, though slightly asymmetric.  The
C$^{18}$O line appears to bisect the CS and HCO$^+$ profiles in velocity
indicating that the two latter lines are likely optically thick and
self-reversed.  The blue-shifted sense of the self-absorption in the self-reversals is
indicative of expansion motions in the outer cloud layers.  Individual spectra of
the HCO$^+$ and \coI\ emission from across \fest\ are simultaneously displayed in
Figure \ref{c18ohcomap} for comparison.  Examination of the figure shows similar
double-peaked, self-reversed HCO$^+$ emission profiles at several positions
around the center of the globule.  In all cases the sense of the asymmetry is the
same suggesting expansion of the outer cloud layers.  In most other positions
the HCO$^+$ lines display blue-shifted self-absorption relative to C$^{18}$O,
also consistent with expansion motions.

\begin{figure}
\begin{center}
\includegraphics[height=4in]{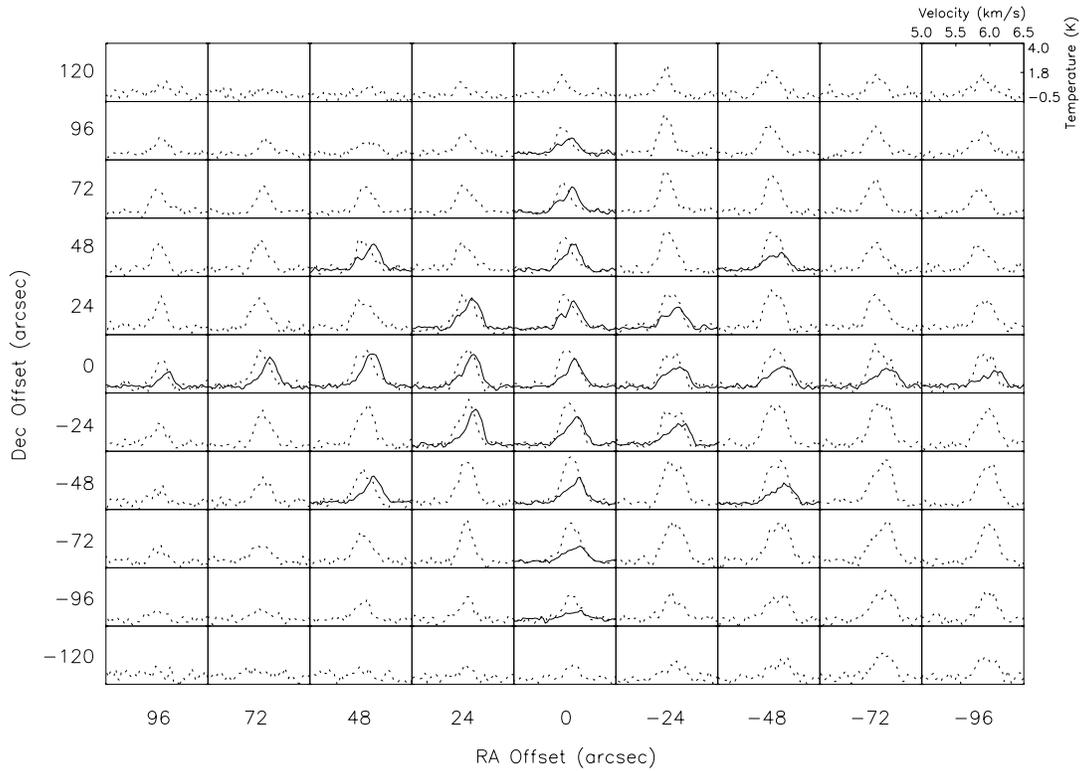}
\caption{Map of \csII\ and \coI\ emission-line spectra from \fest.  
Solid lines are CS emission spectra and dotted
lines are those of C$^{18}$O.  The map is centered at $\alpha_{2000}$ =
17$^h$35$^m$47.5$^s$, $\delta_{2000}$ =
-25$\degr$33$\arcmin$2.0$\arcsec$.  Asymmetric, self-absorbed CS profiles
indicative of expansion are evident across the map}
\label{c18ocsmap}
\end{center}
\end{figure}

As can be seen in Figure \ref{c18ocsmap} CS spectra are also clearly self-reversed
with blue-shifted self-absorption similar to the HCO$^+$ lines in positions where
the HCO$^+$ was also self-reversed, but the signals have lower intensity.  Similar
to HCO$^+$, the other CS spectra appear to exhibit asymmetric profiles relative to
C$^{18}$O with a sense indicative of expansion motions.  The critical density ($3
\times 10^5$ cm$^{-3}$) at which CS emission occurs is the same as that for
HCO$^+$ emission (Ungerechts et al.  1997) so it is reasonable to assume that the
two emission lines probe the same layer of material in the cloud.

A rough estimate of the expansion speed of material was determined by comparing
the velocity of the stronger \hco\ peak to the peak velocity of a \co\ spectrum
at the same position.  The peak velocities were obtained by fitting Gaussian line
profiles to the spectral lines.  This resulted in velocity differences
\mbox{$\delta V$ = $v_{pk}$(\co) - $v_{pk}$(\hco)} which are all negative
confirming that the blueshifted profiles are characteristic of expansion, and
indicating that material is expanding in the cloud over the whole layer under
observation with a mean expansion velocity of -0.09 $\pm$ 0.04~km~s$^{-1}$.  The
same process was applied to the CS spectra and the velocity differences
\mbox{($\delta V$ = $v_{pk}$(\co) - $v_{pk}$(CS))} were also found to be negative
everywhere with a mean difference (expansion velocity) of -0.12 $\pm$ 0.02
km~s$^{-1}$.  This is similar to the range of $\delta V$ for \hco.  This
expanding motion is therefore also evident in the CS line emission profiles.

Another estimate of the expansion speed of material within the cloud
was obtained using the model of \citet{myer96}.  This model can only
be applied to positions (10 positions) where there is a clear double
peak.  In this model, the expansion velocity, $v_{exp}$ is given by,
\begin{equation}
v_{exp} = \frac{\sigma^2}{v_R - v_B}\ 
\ln \frac{1 + e (T_{BD}/T_D)}{1 + e (T_{RD}/T_D)} ,
\end{equation}
where $T_D$ is the brightness temperature of the dip, $T_{BD}$ is the
height of the blue peak above the dip with its corresponding velocity,
$v_B$, $T_{RD}$ is the height of the red peak above the dip with its
corresponding velocity, $v_R$, $\sigma$ is the velocity dispersion of
an optically thin line (\co\ here). For the \hco\ lines, the mean
expansion speed was calculated to be -0.07 $\pm$ 0.02 and
and for the CS lines, -0.07 $\pm$ 0.02 km s$^{-1}$; both 
these estimates are somewhat lower than
those derived from the peak velocity difference method in the previous
paragraph. Nonetheless, though somewhat uncertain, the overall expansion 
speeds we estimate are formally less than the one dimensional sound speed of 
$a \sim$ 0.19 km s$^{-1}$ in a 10 K gas.

\subsection{Velocity Field}\label{sec:vel}

\begin{figure}
\begin{center}
\includegraphics[height=3in]{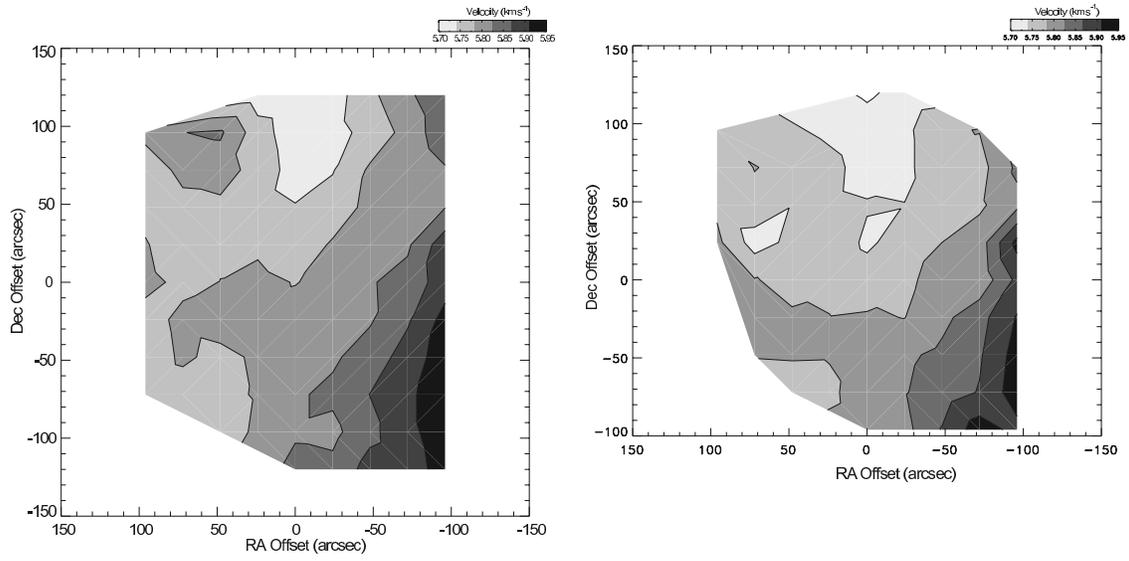}
\caption{\textit{Left.} Peak velocity distribution for \co~($J =
1 \to 0$) emission spectra.
  \textit{Right.}  Peak velocity distribution for \co~($J = 2
\to 1$) emission spectra.}
\label{c18opkv}
\end{center}
\end{figure}

Figure~\ref{c18opkv} shows maps of the variation of the velocity of
the peak in the \co\ spectra.  The peak velocities were determined
from Gaussian fits to the line profiles.  We note that a few of the
\coI\ and \coII\ lines are flat-topped, broadened or slightly asymmetric,
indicating that the lines are slightly optically thick.  The two maps,
however, do reveal a systematic velocity gradient increasing from
upper left to lower right of the map.

To estimate the magnitude and direction of this gradient, the method
of \citet{good93} was used as adapted by \citet{lada03}.  The velocity
gradient is assumed to be linear when projected against the plane of
the sky so the observed velocity $v_{lsr}$ can be related to the
velocity gradient $\ud v/ \ud s$ using
\begin{equation}
v_{lsr} = v_0 + \frac{\ud v}{\ud s}\Delta\alpha~\mathrm{cos} \theta +
\frac{\ud v}{\ud s} \Delta\delta~\mathrm{sin} \theta
\end{equation}
$\Delta\alpha$ and $\Delta\delta$ are Right Ascension and Declination
offsets in arcseconds.  $v_0$ is the systemic velocity of the cloud
and $\theta$ is the angle between north and the 
direction of the velocity gradient of magnitude
$\ud v/ \ud s$.  A least-squares fit of a two-dimensional plane to the
observed $v_{lsr}$ (line center velocity) measurements of \coI, \coII\
and \nhI\ (the \nhI\ spectra contained hyperfine components, and so
the isolated component at the same $v_{lsr}$ as the \coI\ line was
used in this analysis) provided estimates given in Table
\ref{tb:velgradfit}.  The errors for the N$_2$H$^+ $ are larger
because there were only nine spectra measured that were useful for
this purpose.  The \nh\ emission was less extended than the \co\
emission, therefore, more densely sampled observations are needed
to confirm the estimates using the N$_2$H$^+ $ data.

If the velocity gradient derived previously is removed, the average
velocity difference between the neighboring spectra is then
essentially zero ($\sim 0.0025$ km s$^{-1}$).

\subsection{Line Width Distribution}\label{sec:linwid}

\begin{figure}
\begin{center}
\includegraphics[height=3.5in]{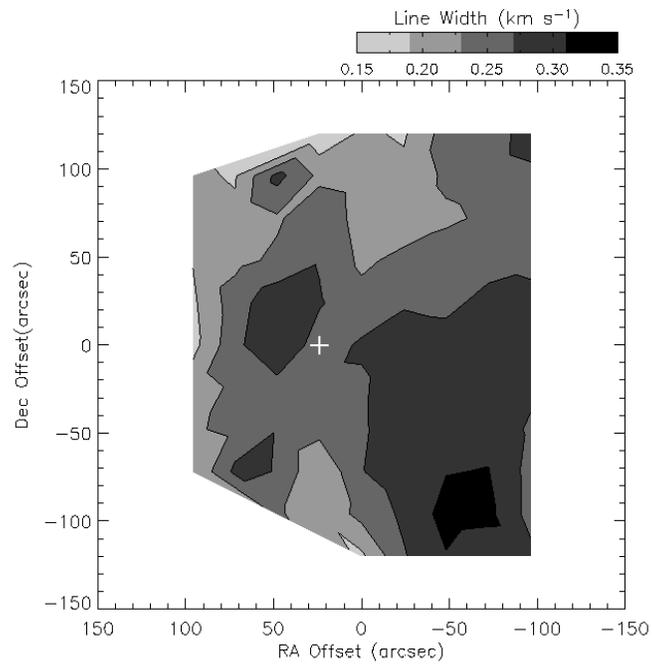}
\caption{Line width distribution of \coI\
spectra.  The white cross denotes the position of the dust extinction
peak.}
\label{c18olnw}
\end{center}
\end{figure}

Line widths (FWHP) were determined from Gaussian fits to the observed \coI\ and
\coII\ line profiles.  The \nhI\ spectra have several hyperfine components so
the line widths were determined from a Gaussian fit to the isolated ($F_1 =
0-1$) component.  The resulting \coI\ and \coII\ line widths were found to be
in a range from $\sim$~0.19 to $\sim$~0.35~km~s$^{-1}$.  The \nh\ line widths
were narrower, and were in the range $\sim$ 0.15 to $\sim$~0.25 km s$^{-1}$.
Figure~\ref{c18olnw} shows the variation in line width of the \coI\ line
profiles.  Because some of these lines are optically thick (see comment in
\S~\ref{sec:vel}), the line widths determined from a single Gaussian fit are
slightly larger than they would be if the lines were optically thin.
Nevertheless, the line widths seem to increase slightly at positions away from
the highly extincted region around offset (24$\arcsec$,~0$\arcsec$), marked with
a white cross in Figure~\ref{c18olnw}.  This is similar to B68 where the
linewidths also appeared to increase away from its center.  The reason for this
is not clear.

No independent measure of the kinetic temperature of
\fest\ has as yet been made; a value of 10~K has therefore been
adopted in this paper because data has shown that this applies to most
globules and dense cores (e.g., \citet{bens89}).  The thermal line
width, $\surd[(8\mathrm{ln}2) kT_K/(\mu m_H)]$ for \coI\
lines was therefore calculated to be 0.123~km~s$^{-1}$, and for \nh
lines, 0.125~km~s$^{-1}$.  The nonthermal contribution to the line
width was then estimated using the relation,
\begin{equation}
(\Delta v_{obs})^2 = (\Delta v_{th})^2 + (\Delta v_{nth})^2 ,
\end{equation}
where $\Delta v_{obs}$ is the observed line width, $\Delta v_{th}$ is
the thermal line width and $\Delta v_{nth}$ is the nonthermal line
width.  The resulting average nonthermal line width for the \coI\
lines was $\sim$~0.25~km~s$^{-1}$, and for the \nh\ lines,
$\sim$~0.15~km~s$^{-1}$. 
To the extent that these lines are optically
thin, this suggests that in producing the observed line profiles,
nonthermal broadening mechanisms, e.g., turbulence, may play
a role. There may be more turbulent motion in the outer layers probed
by the \co\ molecular lines than in the inner more dense layers probed
by the \nh\ line. However, the corresponding one dimensional non-thermal
velocity dispersions ($\sigma_{nth}$) are 0.11 and 0.06 km~s$^{-1}$
for the \co\ and \nh\ emitting gas, respectively. These values are
both subsonic, that is, less than the one dimensional 
sound speed (0.19 km~s$^{-1}$) in a 10 K gas.

\subsection{Spatial Distributions of Molecular Emission and Dust Extinction}

In Figure 6 we show contour maps of C$^{18}$O (1--0), N$_2$H$^+$ (1--0), and
HCO$^+$ (1--0) integrated emission overlaid on a grey-scale map of the
distribution of visual extinction. The extinction map was constructed 
from the data of Alves et al. (2002) and convolved
with a 24 arc sec Gaussian smoothing kernel to match the resolution
of the molecular-line data.  The patterns seen in this figure are comparable to those seen in
numerous other low-mass pre-stellar cores such as B68 or L1544 (Bergin et al.
2002; Tafalla et al.  2002).  In particular, both C$^{18}$O and HCO$^+$ show
emission distributions that are broader than the distribution in extinction with peaks well
separated from the extinction maximum.  In contrast N$_2$H$^+$ shows the highest
degree of agreement with the dust extinction.  This pattern is attributed to the
freeze-out of molecules on the surfaces of cold dust grains in gas where the
density exceeds 10$^5$ cm$^{-3}$ (e.g.  Bergin \& Langer 1997; Aikawa et al.
2005).  In this scenario the freeze-out of CO and its isotopologues leads to the
formation of N$_2$H$^+$ in the dense gas dominated by freeze-out and traced by
the dust.  HCO$^{+}$ has structure similar to CO which is not surprising since
HCO$^+$ likely forms from gas phase reactions involving CO.  For a
summary of this process in starless cores see Bergin \& Tafalla (2007).

\section{Discussion}\label{sec:disc}

\subsection{Is FeSt 1-457 Gravitationally Bound?}\label{sec:grav}

One explanation for the expansion of material from \fest\ could be that the
globule is simply unbound.  This might be expected if the core is a transitory
feature in a global turbulent sea \citep{ball06}.  In this section, we assess
this possibility by determining whether or not the core is bound.  A mass for
\fest\ of $\sim 3.1\ [\frac{d(pc)}{130}]^2$ $M_{\sun}$ was derived by spatially
integrating the extinction data over the area of the globule, after correcting
for a significant foreground/background extinction component (A$_V$ $\sim$ 6
magnitudes) and an assumed distance of $d$~pc. The magnitude of the background
extinction was derived from examination of the azmuthially averaged extinction
profile of the globule constructed from the Alves et al. (2002) deep extinction
data and is the same as that derived by Alves et al.  (2007) for this core
from a wavelet decomposition of their 2MASS extinction map of the Pipe cloud.
The escape velocity ($\surd [2GM/R]$) from \fest\ is estimated to be $\sim 0.6$\
$[\frac{d(pc)}{130}]^{0.5}$ km s$^{-1}$.  The average three-dimensional velocity
dispersion ($\sqrt{3a^2 + 3\sigma^2_{nth}}$) of the bulk gas in the globule
is $\sim$ 0.3-0.4~km~s$^{-1}$, and  significantly
less than the escape velocity. Thus the globule is likely to be gravitationally
bound for the assumed distance of 130 pc or for any other plausible distance to
the cloud.  Moreover, the expansion speeds detected in the self-absorbed
molecular line data ($\sim$ 0.1 km s$^{-1}$, see \S~\ref{sec:hco}) are also
significantly less than the calculated escape speed.  The expansion of the outer
layers of the globlule appears not to be caused simply by evaporation of its gas
as in an unbound, transient cloud.

 A Jeans mass of $3.8~$M$_{\sun}$ was derived using $M_J =
18~$M$_{\sun}~T_K~^{1.5}~\bar{n}^{-0.5}$ where $T$ is the globule's kinetic
temperature assumed to be $\sim$ 10~K.  However, if we assume the lines are only
thermally broadened, with no turbulence, then the kinetic temperature is 17~K and
this doubles the Jeans mass estimate (\S~\ref{sec:linwid}).  The average density
$\bar{n} $=$ 2.2\times 10^4$ cm$^{-3}$ was determined using the extinction data
and a radius, $R = 2.6 \times 10^{17}$ cm.  The radius was determined to be the
point at which the radially averaged extinction profile reaches the background and
marks the outer edge of the globule.  Since the mass of \fest\ is comparable to
its Jeans mass, we conclude that it is likely marginally stable against
gravitational collapse.  If thermal pressure is not the only source of internal
support then the globule could be even more stable against collapse.  More
detailed analysis of the globule's structure would be required to better evaluate
its overall stability.

\subsection{Possible Rotation?}

There may be some contribution to stability of the globule from rotation.
Assuming solid body rotation, $\beta$, the ratio of rotational kinetic energy to
gravitational energy, is given by \citep{good93},

\begin{equation}
\beta = \frac{(1/2)I \omega^2}{q G M^2 /R} =  \frac{1}{2} \frac{p}{q} \frac{\omega^2 R^3}{G M}
\end{equation}

\noindent
$R$ is the globule's radius, $M$ its mass and $p$ is defined such that
the moment of inertia, $I = p\hspace{0.1em}MR^2$, and $q$ is defined
such that the gravitational potential energy is
$q\hspace{0.2em}GM^2/R$.  $\beta$ was estimated to be 0.01 for
\fest\ using $M = 3.1~$M$_{\sun}, R = 2.6 \times 10^{17}$ cm,
$\omega = (\ud v/\ud s)/\sin i = $ [1.5~km~s$^{-1}$~pc$^{-1}$]/$\sin
i$ (for \coI, see Table~\ref{tb:velgradfit}),
where $i$ is the inclination of the globule to the line of sight, and assuming
$p/q = 0.22$ as for a sphere with a density profile $r^{-2}$ and
$\sin i =1$.  The contribution of rotation to the overall stability of
\fest\ is therefore not significant.  Note that $\beta$ of 0.01 is also
consistent with the results of \citet{good93} that most clouds have
$\beta \leq 0.05$.

\subsection{Thermal vs. Turbulent Support}\label{sec:therm}

Thermal motions may be a significant source of pressure support for \fest.  This
can be quantified by evaluating the ratio of thermal to nonthermal (turbulent)
pressure given by, 

\begin{equation} 
R_p = \frac{a^2}{\sigma_{nth}^2} ,
\end{equation} 
\noindent

where $a$ is the one dimensional isothermal sound speed and $\sigma_{nth}$ is
the one dimensional nonthermal or turbulent velocity dispersion and
$\sigma_{nth} = [\Delta v_{nth} / \surd[8\ln2]]$. Assuming a gas temperature
of 10 K, the average value of $R_p$ for all the
C$^{18}$O observations is 3.75 $\pm$ 1.95, which may be an underestimate for
those lines that may be somewhat thick.  The average value of $R_p$ for the \nh\
data is 6.09 $\pm$ 2.07.  These values indicate that the thermal pressure
significantly exceeds the turbulent pressure in the globule, and absent strong
static magnetic fields, is the dominant source of internal support against
gravity.

In comparing the turbulent velocity dispersion to the isothermal sound
speed of 0.19~km~s$^{-1}$ in a 10~K molecular gas, the nonthermal
motions appear to be subsonic over the surface of \fest.  If the
\nh\ observations probe a deeper layer of the globule than the
\co\ lines, these motions are even more subsonic in the inner layers of
the globule. These considerations further indicate that thermal motions  
provide a significant source of internal pressure.

The apparent velocity gradient in the \co\ data was calculated and removed in
\S~\ref{sec:vel}.  The resulting average velocity difference between neighboring
spectra was essentially zero (\mbox{$\sim 0.0025$ km s$^{-1}$}) so there appears
to be no residual scatter in peak velocities of the spectra on scale sizes
comparable to the beam width that could be caused by turbulent motions.  This
also is consistent with significant thermal support for the globule.

\subsection{Pulsating Globule?}

In the absence of an embedded source in \fest\ that could cause outflow of
material, it is difficult to explain the observations reported here.  In the
previous section we showed that the globule is gravitationally bound with thermal
pressure as the dominant source of internal support.  But what causes the observed
expansion motions?   The facts that 1- the globule is bound, 2- thermally
supported and 3- does not exceed the Jean's mass, suggest that this core is stable
and not far from a state of hydrostatic equilibrium.  Moreover, the expected
(one-dimensional) virial velocity dispersion, $ \sigma_{virial} = \surd[{1\over
5}GM/R]$, is 0.18 km s$^{-1}$ and is comparable to the sound speed in a 10 K gas
as would be expected for a thermally dominated, stable core.  Indeed, we derive
the velocity dispersion for the H$_2$ gas in the core to be $\sigma = \sqrt{a^2 +
\sigma_{nth}^2} \approx $ 0.21 km s$^{-1}$ close to the predicted value.  However,
given that its outer layers are globally expanding, the globule cannot be
presently in a strict equilibrium state.  One viable explanation that can
simultaneously account for all these facts, is that \fest\ is in dynamical
oscillation around an equilibrium state.  The outflowing motion is then part of
a mode of oscillation such that its layers pulse in and out with a period of
possibly a few hundred thousand years.  We further explore this
interesting possibility below.

In their study of molecular lines observed in the globule B68, \citet{lada03}
concluded that that globule could be experiencing low order mode, nonradial
oscillations of its outer layers.  They suggest that the globule may have
interacted with the shock of a supernova remnant, which instigated a
perturbation that set at least its outer layers in oscillatory motion.  Figure~7
of their paper displays the real parts of the spherical harmonic functions for a
range of low order modes for an oscillating sphere.  The $l = 2, m = 2$ mode
corresponds closely to the pattern observed in B68.  A study by \citet{keto06}
demonstrated that linear perturbations on a pressure-bounded thermally supported
sphere could indeed produce the spatial variation in profile shape observed in
B68.  \fest\ could also be oscillating in this mode but the geometry is such
that we are looking `edge-on' to the pulsating cloud.  This means the mode of
pulsation appears similar to the `breathing' or $l = 0$ mode, i.e., outflow
motion over the whole surface of the cloud \citep{keto06}.

Calculations were carried out to find the modes of pulsation for an
isothermal, spherical globule of gas with similar parameters (e.g.,
radius = $2.6 \times 10^{17}$ cm, density = $2.2\times 10^4$
cm$^{-3}$ , internal sound speed = 0.18~km~s$^{-1}$) as
for \fest.  Small wave-like perturbations were introduced, and a
differential wave equation was determined using the perturbed and
unperturbed forms of its equations of state.  The modes of oscillation
postulated are likely to be acoustic, with relatively large amplitudes
and energies, so that the linear approximation that we use for the mode
frequencies is a simplification of the full dynamical problem. In this
linear approximation, we ignore the gravitational modes, and
find that the acoustic modes follow the dispersion relation 

\begin{equation}\label{eq:disp1}
\mathfrak{D}(\omega) =
\frac{h_l^{(1) \prime}(x_{\omega})}{ h_l^{(1)} (x_{\omega})} -
\frac{c_{out}}{c_{in}} \ \frac{\rho_{0(out)}}{\rho_{0(in)}} \ 
\frac{j_l^{\prime} \left( x_{\omega} \left[c_{out}/c_{in}
\right] \right) }{j_l \left( x_{\omega} \left[ c_{out}/c_{in} \right] \right)} = 0
\end{equation}

\noindent
$x_{\omega} = \omega R_0 / c_{out}$ where $\omega$ is the frequency of
the oscillations; $R_0$ is the unperturbed radius; $c_{in}, c_{out}$
are isothermal sound speeds inside and outside the globule
respectively; $\rho_{0(in)}, \rho_{0(out)}$ are the unperturbed
densities inside and outside the globule respectively; and,
$h_l^{(1)}$ and $j_l$ are spherical Hankel and Bessel functions 
of the first kind or order $l$, with the prime denoting differentiation with
respect to the argument.

The frequency $\omega$ is complex and roots of the dispersion relation where
$\omega$ has a negative imaginary part, correspond to decaying modes of
oscillation.  The required mode for pulsation is the least-damped mode or the
mode which decays at the slowest rate; this is the mode with the smallest
negative imaginary part of $\omega$.  This mode was calculated to be $l = 2$
with a frequency of $\sim 9 \times 10^{-14}$ Hz corresponding to an oscillation
period of $\sim 3 \times 10^5$~years, comparable to the sound crossing time.  It
is plausible that this oscillation was excited by some transient dynamical
event, such as a shock, that deposits an energy greater than about $3 \times
10^{43}$~ergs, (our best estimate of the kinetic energy in coherent motions)
into the outer regions of \fest, where \hco\ molecules exhibit a coherent
pattern of expansion motions.  Calculations by \citet{keto06} show that such
large amplitude motions (i.e., motions comparable to the sound speed) can cause
significant geometrical distortions of a core.  An important point to note is
that in the high resolution extinction map of \fest\ obtained by \citet{alve02},
a slight splitting of the core's central regions is visible.  This splitting
could be another manifestation of an $l = 2$ mode of oscillation.

\fest\ is situated in the direction of the Pipe Nebula, which lies at the edge
of the Scorpio Centaurus OB Association.  \citet{onis99} suggested that stellar
wind from the B0 type star $\tau$~Sco caused compression of the molecular gas,
triggering star formation in the B59 cloud, situated $\sim$~5$\degr$ west and
$\sim$~3$\degr$ north of \fest.  In Figure~5 of their paper, \citet{onis99} show
an observation of \fest, designating it `Core 12'.  In discussing the effect of
the OB stars on the Pipe Nebula, \citet{onis99} estimated that $\sim 1 \times
10^{46}$ ergs has been deposited in the cloud over a period of $1 \times 10^7$
years.  If \fest\ is indeed near this OB association and was also able to
intercept only 0.3\% of the estimated energy deposition, then the effects of the
postulated shock waves may be more widespread than previously speculated.  Not
only did they trigger star formation in B59, but they may have also set gaseous
globules such as \fest\ and B68 (which is also part of the Pipe complex) into
oscillation.  More observations and comparison with theoretical models of cloud
pulsation are needed to confirm this hypothesis.

\subsection{Molecular Depletion}

In Figure 7 we provide a direct comparison of the dependence of C$^{18}$O (1--0)
and N$_2$H$^+$ (1--0) emission with visual extinction at the same angular
resolution.  For both species the figures display trends that are similar to
those seen in B68 by Bergin et al.  (2002).  In these plots, a linear
correlation between integrated intensity and A$_V$ corresponds to a constant
abundance throughout the cloud.

%
%
%

The C$^{18}$O (1--0) emission shows
a linear dependence with A$_V$ until $\sim 10-12$ mag whereupon the trend
flattens.  
Given the drop in the 
the C$^{18}$O integrated emission/A$_V$ relation near A$_V \sim 12$ mag we 
have fit the following function: 
$\int T_{mb}dv ({\rm C^{18}O}) = a + b(A_V - 12)$ to the data.  
We  find an intercept of $a = 1.09\pm 0.01$ K km s$^{-1}$ and a slope of
\begin{equation}
b= 
\begin{cases} 0.117\pm0.002 & 
\text{if $A_V \le 12^m$,}
\\
0.002\pm0.001 &\text{if $A_V > 12^m$.}
\end{cases}
\end{equation}

\noindent 
Thus for A$_V <$ 12$^m$, where we see a significant linear correlation between
gas emission and dust extinction, we estimate a C$^{18}$O abundance of $\sim
10^{-7}$ (relative to H$_2$).  This is comparable to that measured previously by
Frerking, Langer, \& Wilson (1982) of $1.7 \times 10^{-7}$ in Taurus and to that
obtained from a similar comparison of extinction and CO measurements toward L 977
(Alves et al.  1999).  For A$_V > 12^m$ the correlation is essentially flat,
indicative of saturation or, as in B68, a drop in abundance towards denser
regions.  We can use the intercept to estimate average abundance in this gas and
find that the abundance is $\sim (2 - 4) \times 10^{-8}$ between 20 and 40
magnitudes.  Thus, we estimate that CO is depleted by a factor of $\sim$5.  In \S
3.2 we noted that the C$^{18}$O emission lines show indications of being slightly
optically thick.  Thus these abundance estimates 
are upper limits to the {\em average} depletion in
the cloud.

The situation for N$_2$H$^+$ (1--0) in Figure 7 is more complicated.  There is a
rough correspondence between integrated emission and A$_V$ and it is well fit by a
line with a slope of 0.106$\pm$0.001 K km s$^{-1}$ mag$^{-1}$ and an intercept of
-0.711$\pm$0.031 K km s$^{-1}$.  This is consistent with an abundance of
10$^{-10}$ relative to hydrogen.  However, we cannot exclude the possibility that
the trend becomes flat at $A_V > 20$~mag.  Fits to the intensities of the
hyperfine lines of \nhI\ suggest that the total opacities are of order 12 at the
emission peaks, so that all the hyperfine components are saturated.  This is not
the case at low A$_V$, where the lines appear to be optically thin.  Thus we
believe that the line integrated intensity-$A_V$ correlation may be turning over
at high A$_V$ as a result of saturation and rather than depletion. However, we
also note that the spatial map of N$_2$H$^+$ in Figure 6 displays peaks northeast
and southwest of the dust extinction peak and this could also be an indication of
depletion of this species similar to what has been observed in B68 (Bergin et al.
2002).  However, due to the high opacity of the line, it is not possible to
separate the effects of spatial variations in excitation from those of column
density without detailed chemical modeling.

\section{Summary}

High resolution radio observations were made of the globule \fest.
Molecular lines observed were \csII, \hcoI, \coI, \coII\
and \nhI.  The \hco\ and CS spectra showed clear self-reversed,
asymmetric profiles across the face of the globule.  The sense of the
asymmetry is indicative of the presence of global expansion motions
across the globule.  These motions appear to be 
subsonic and significantly below the
escape speed of the cloud.  A search for IRAS, MSX sources and Spitzer
sources showed the globule to be starless, and therefore the origins
of these expansion motions is unclear.  

In this paper, we propose the explanation that the globule is in a
quasi-stable state near hydrodynamic equilibrium such that its outer
layers are pulsating.  It is possible that a passing shock wave from a
nearby OB star association has excited the globule setting its outer
layers into motion.  Such oscillatory motion is visible in
the layers on the Sun (where the fundamental radial period is $\sim$ 1
hour and other oscillation periods are $\sim$ 5 minutes) but in \fest,
the periods are thought to be of the order of hundreds of thousands of
years.  This is similar to what \citet{lada03} observed in B68.
Calculations to determine the dominant mode of pulsation resulted in
an $l = 2$ mode with a frequency of $\sim 9 \times 10^{-14}$ Hz, and
an oscillation period of $\sim 3 \times 10^5$ years. A high resolution 
extinction map of \fest\ exhibits a double peak structure at
the highest levels of opacity, corresponding to visual extinctions
of $\sim$50 mag \citep{alve02}.  It is possible that the proposed 
$l = 2$ mode of oscillation could explain the observed splitting of the 
core in this column density map.

Finally, we find evidence for significant depletion of CO and perhaps 
\hco\ in this globule. However, \nhI\ may be undepleted to 
depths of about 40 magnitudes of visual extinction in the core of
the cloud.



\acknowledgments

We are grateful to Dr.  Tracy Huard and the staff at the IRAM 30 m telescope for
their assistance with the observations.  We thank Dr.  Carlos Roman-Zuniga for
assistance in constructing the appropriate extinction map and with calculating the
cloud mass.  We thank the referee for insightful suggestions that strengthened the
presentation.  EDA is particularly indebted to Dr.  Mike Masheder for his able
guidance, useful suggestions and continued support throughout all aspects of this
project.  EDA was supported by a PPARC Postgraduate Studentship.  CJL acknowledges
support from NASA Origins grant NAG-13041.


\clearpage


\clearpage

\clearpage

\clearpage

\clearpage

\clearpage

\begin{figure}
\begin{center}
\includegraphics[angle=270,width=\textwidth]{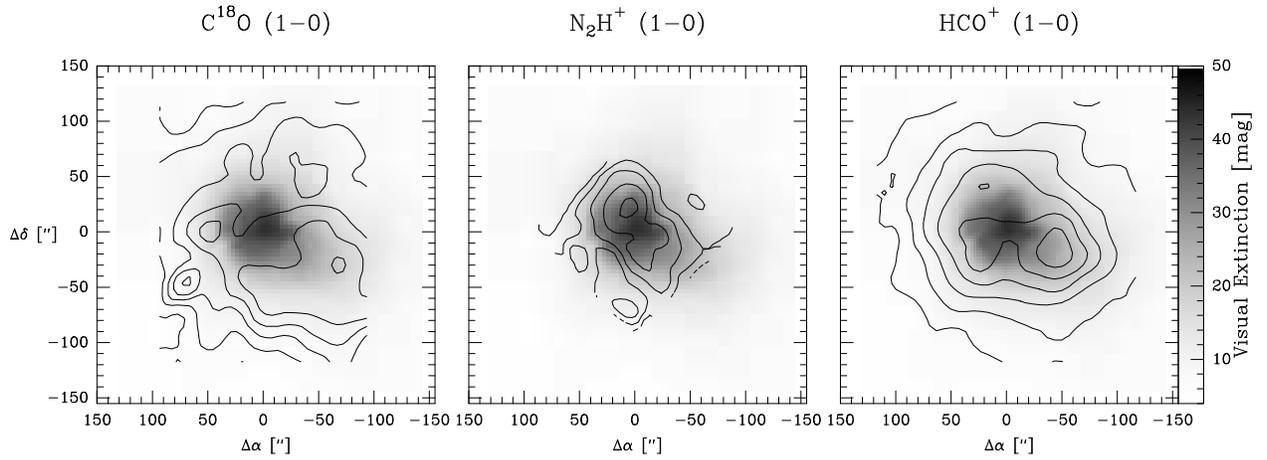}
\caption{
Comparison of the spatial distributions of molecular emission shown as 
contours with a map of 
visual extinction constructed with the same angular resolution and shown as 
grey scale.  Contour levels are given in 
$\int T_A^*\;dv$ with units of K km s$^{-1}$ (C$^{18}$O: 0.2 to 2.0 by 0.2; 
N$_2$H$^+$: 0.1 to 0.8 by 0.1;  HCO$^{+}$: 0.2 to 1.2 by 0.2). 
The extinction scale is shown in the bar on the right of the figure and
the extinction ranges from approximately 6 - 43 magnitudes. For clarity
the grey-scale contrast
has been adjusted to emphasize the dust distribution in the center of the core.
}
\label{festcompare}
\end{center}
\end{figure}

\clearpage

\begin{figure}
\begin{center}
\includegraphics[width=\textwidth]{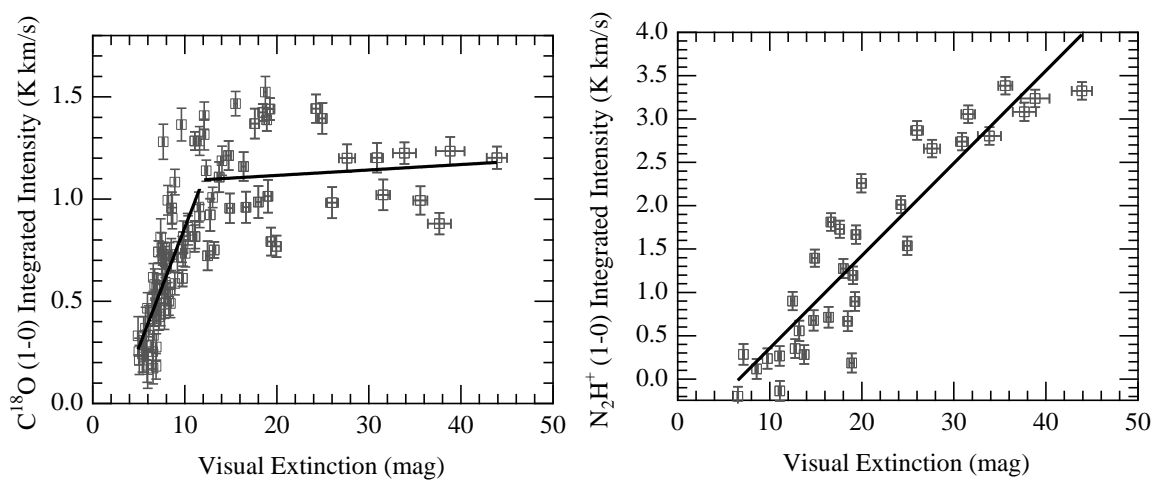}
\caption{
({\em Left}) C$^{18}$O J=1-0 integrated intensity as a function of 
visual extinction for
the entire FeSt 1-457 dark cloud. ({\em Right})  N$_2$H$^+$ J=1-0 integrated 
emission as a function
of visual extinction.  In all plots the data are presented as open
squares with error bars while solid traces are the result of linear fits to 
the data. The extinction and molecular line observations in both 
plots were obtained at the same angular resolution.
}
\label{avplots}
\end{center}
\end{figure}


\clearpage

\begin{deluxetable}{lcccc}
\tablewidth{0pt}
\setlength{\tabcolsep}{3mm}
\tablecaption{Line parameters. \label{tb:linpars}}
\tablehead{
\colhead{Line} &
\colhead{$\nu$\tablenotemark{a}} &
\colhead{HPBW\tablenotemark{b}} &
\colhead{$\Delta v$\tablenotemark{c}} \\
\colhead{} & \colhead{(GHz)} & \colhead{($\arcsec$)} & \colhead{(km s$^{-1}$)} }
\startdata
\csII & 97.980968 & 25 & 0.030 \\
\hcoI & 89.188512 & 28 & 0.033 \\
\coI & 109.78218 & 22 & 0.027 \\
\coII & 219.560319 & 11 & 0.013 \\
\nhI &  93.173178 & 26 & 0.031 \\
\enddata
\tablenotetext{a}{{Line rest frequency}}
\tablenotetext{b}{{Half power beam width}}
\tablenotetext{c}{{Velocity resolution}}
\end{deluxetable}


\clearpage

\begin{deluxetable}{lcccc}
\tablewidth{0pt} \setlength{\tabcolsep}{3mm} 
\tablecaption{Results of velocity gradient fit.\tablenotemark{a} \label{tb:velgradfit}}
\tablehead{ \colhead{Line} & \colhead{$v_0$\tablenotemark{b}} &
\colhead{$\ud v / \ud s$\tablenotemark{c}} & \colhead{$\ud v / \ud
s $ at 160 pc} & \colhead{$\theta$\tablenotemark{d}} \\ 

\colhead{} & \colhead{(km s$^{-1}$)} & \colhead{(m s$^{-1}$ arcsec$^{-1}$)} &
\colhead{(km s$^{-1}$ pc$^{-1}$)} & \colhead{($\degr$)} } 

\startdata
\coI & $5.81 \pm 0.001$ & $0.73 \pm 0.012$ & 1.5 & $231.5 \pm 1.0$ \\

\coII & $5.79 \pm 0.001$ & $0.70 \pm 0.022$ &
 1.4 & $238.9 \pm 1.9$ \\ 

\nhI & $5.79 \pm 0.003$ & $1.13 \pm 0.13$ &
 2.3 & $249.8 \pm 7.5$ \\
\enddata

\tablenotetext{a}{{Errors quoted are $1 \sigma$ uncertainty}}
\tablenotetext{b}{{Systemic velocity}}
\tablenotetext{c}{{Magnitude of velocity gradient}}
\tablenotetext{d}{{Direction of velocity gradient measured East of North}}
\end{deluxetable}

\clearpage


\end{document}